\documentclass[%
 reprint,
superscriptaddress,
 amsmath,amssymb,
 aps,
]{revtex4-1}

\usepackage{graphicx}% Include figure files
\usepackage{dcolumn}% Align table columns on decimal point
\usepackage{bm}% bold math
\usepackage{booktabs} %需要加载宏包{booktabs}
\usepackage{float}
\usepackage{subfigure}
\usepackage{color}

\begin{document}

\preprint{APS/123-QED}

\title{Twin-field quantum key distribution with passive-decoy state}% Force line breaks with \\

\author{Jun Teng}
\altaffiliation[]{These authors contribute equally to this work.}
\author{Feng-Yu Lu}
\altaffiliation[]{These authors contribute equally to this work.}
	\author{Zhen-Qiang Yin}
	\email{yinzq@ustc.edu.cn}
	\author{Guan-Jie Fan-Yuan}
	\author{Rong Wang}
	\author{Shuang Wang}
	\author{Wei Chen}
	\affiliation{CAS Key Laboratory of Quantum Information, University of Science and Technology of China, Hefei 230026, P. R. China}
	\affiliation{Synergetic Innovation Center of Quantum Information $\&$ Quantum Physics, University of Science and Technology of China, Hefei, Anhui 230026, P. R. China}
	\affiliation{State Key Laboratory of Cryptography, P. O. Box 5159, Beijing 100878, P. R. China}
	\author{Wei Huang}
	\author{Bing-Jie Xu}
	\affiliation{Science and Technology on Communication Security Laboratory,
		Institute of Southwestern Communication, Chengdu, Sichuan 610041, China}
	\author{Guang-Can Guo}
	\author{Zheng-Fu Han}
	\affiliation{CAS Key Laboratory of Quantum Information, University of Science and Technology of China, Hefei 230026, P. R. China}
	\affiliation{Synergetic Innovation Center of Quantum Information $\&$ Quantum Physics, University of Science and Technology of China, Hefei, Anhui 230026, P. R. China}
	\affiliation{State Key Laboratory of Cryptography, P. O. Box 5159, Beijing 100878, P. R. China}

\date{\today}% It is always \today, today,

\begin{abstract}
Twin-Field quantum key distribution (TF-QKD) and its variants, e.g. Phase-Matching QKD, Sending-or-not-sending QKD, and No Phase Post-Selection TFQKD promise high key rates at long distance to beat the rate distance limit without a repeater. The security proof of these protocols are based on decoy-state method, which is usually performed by actively modulating a variable optical attenuator together with a random number generator in practical experiments, however, active-decoy schemes like this may lead to side channel and could open a security loophole. To enhance the source security of TF-QKD, in this paper, we propose passive-decoy based TF-QKD, in which we combine TF-QKD with the passive-decoy method. And we present a simulation comparing the key generation rate with that in active-decoy, the result shows our scheme performs as good as active decoy TF-QKD, and our scheme could reach satisfactory secret key rates with just a few photon detectors. This shows our work is meaningful in practice.
\end{abstract}

\pacs{Valid PACS appear here}% PACS, the Physics and Astronomy
\maketitle

\section{\label{sec:level1}INTRODUCTION }

With the help of quantum key distribution (QKD), distant agents (Alice and Bob) are able to share secret keys with information-theoretic security which is guaranteed by quantum physics \cite{bennett1984proceedings,ekert1991quantum,lo1999unconditional}. QKD has been developed rapidly both in theory and experiment over the decades \cite{inoue2002differential,gobby2004quantum,zhao2006experimental,lo2012measurement,wang2015phase,zhou2016making,yin2016measurement,liao2017satellite,wang2020optimized}. However, a fundamental limit is that the secret key rate (SKR) is scaled by the transmittance $\eta$ \cite{takeoka2014fundamental,pirandola2017fundamental,wilde2017converse}, which is finally revealed as the linear key rate bound: $R \leq -\rm log_2(1-\eta)$ {\cite{pirandola2017fundamental}. Fortunately, the Twin-Field quantum key distribution (TF-QKD) protocol made a difference \cite{lucamarini2018overcoming}, in TF-QKD protocol, pairs of phase-randomized optical pulses are generated at Alice and Bob and then combined at a central measuring station Charlie, single-photon interference is produced at Charlie, this lets TF-QKD gain the square-root of the channel transmittance, which is the same as it with a quantum repeater. The Ref. \cite{tamaki2018information} enhanced the security proof of that given in Ref. \cite{lucamarini2018overcoming}, it proposed a novel way to use decoy states by switching between decoy mode and code mode, by doing this, it confirmed the scaling properties of the original scheme and overcame the linear key bound at long distances.

Inspired by the original TF-QKD, many variants are proposed including Phase-matching protocol \cite{ma2018phase}, {Sending-or-not-sending protocol \cite{wang2018twin}. And Ref. \cite{cui2019twin,curty2019simple,lin2018simple} removed active phase randomization and phase postselection in code mode, this proved to reach a higher final key rate. Yet, the security proof of all these protocols are based on decoy-state method \cite{hwang2003quantum,lo2005decoy,wang2005beating}, the essential idea behind decoy-state method in TF-QKD is simple, the sender (Alice and Bob) varies the mean photon number of each pulse she transmits to the central station (Charlie), users can obtain a better estimation of the behavior of the quantum channel from the measurement results corresponding to different intensity settings, this helps to enhance the achievable SKR and communicating distance. In practical experiments, decoy-state method is usually performed by actively modulating a variable optical attenuator together with a random number generator  \cite{schmitt2007experimental}. However, these approaches may lead to side channel and can even break the security of the system \cite{jiang2012wavelength}, if the VOA, which changes the intensity of Alice’s (Bob’s) pulses, is not properly designed, it may happen that some physical parameters (for example frequency spectrum) of the pulses emitted by the sender depend on the particular setting selected, which could open a security loophole in the active schemes. Fortunately, passive-decoy protocol \cite{mauerer2007quantum,adachi2007simple,ma2008quantum} is able to remove certain side channel caused by actively modulating intensity of the source. Specifically, in our proposed scheme, this kind of side channel can be removed provided that Alice and Bob have perfect local detectors and there is no Trojan horse attack. Even though perfect detectors are not available in practical experiments, the side channel may still be significantly reduced when high-detection-efficiency local detectors are equipped. The original passive-decoy protocol utilizes a parametric downconversion source in conjunction with a photon number resolving detector to substitute an idealized single-photon source, the passive-decoy state selection can be accomplished without the need for any active optical elements. Besides, the Refs.\cite{curty2009non,curty2010passive,zhang2018simple} implemented passive-decoy method by using weak coherent pulses (WCPs), in which Alice and Bob treat click and no click of the detectors separately and distill a secret key from both of them. 

\begin{figure*}[htb]
\centering

\subfigure[System of passive-decoy TF-QKD scheme]{
\begin{minipage}[t]{1\textwidth}
\centering
\label{fig:system}
\includegraphics[width=0.9\textwidth]{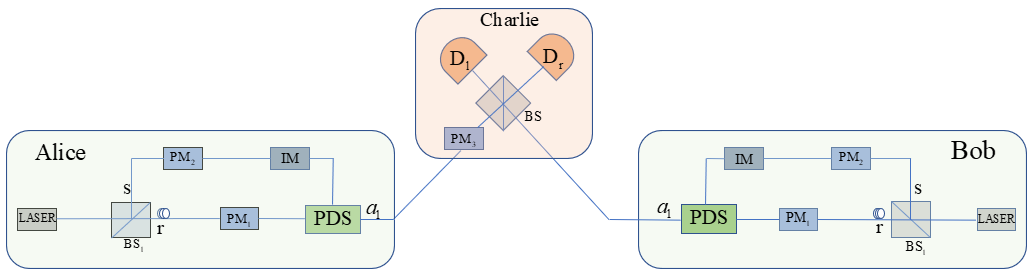}
%\caption{}
\end{minipage}%
}%
 
               %这个回车很重要
\subfigure[Two-intensity setup]{
\begin{minipage}[t]{0.5\textwidth}
%\centering
\includegraphics[width=0.55\textwidth]{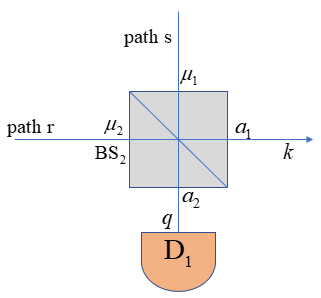}
%\caption{fig2}
\label{fig:intensity_2}
\end{minipage}
}%
\subfigure[Four-intensity setup]{
\begin{minipage}[t]{0.47\textwidth}
%\centering
\includegraphics[width=0.55\textwidth]{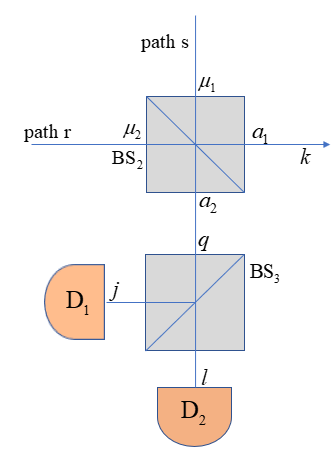}
%\caption{fig2}
\label{fig:intensity_4}
\end{minipage}
}%
\label{fig:my_whole_system}
\centering
\caption{ (a) Schematic diagram of the passive-decoy TF-QKD: PDS, passive-decoy setup; BS, beam splitter; PM, phase modulator; IM, intensity modulator. Alice and Bob have the same setup, $\rm BS_1$ splits the pulse into path s and path r, when code mode is selected, IM blocks the pulse on path s. when decoy mode is selected, $\rm PM_1$ and $\rm PM_2$ randomize the phase of the pulses on each path and leave the IM unmodulated. (b) Two-intensity setup of passive-decoy TF-QKD scheme. $\rm D_1$, single-photon detector. (c) Four-intensity setup of passive-decoy TF-QKD scheme.}
\end{figure*}

To enhance the source security of TF-QKD, in this paper we propose passive-decoy TF-QKD, in our scheme, different decoy states are generated passively according to the mode of photon detectors. Our passive scheme can eliminate the side channel generated by active modulation of source intensities, simultaneously, it provides a performance as good as the active decoy method when the setup with four-intensity settings is applied. In Sec.\ref{sec:level2}, we introduce our scheme and the setups. In Sec.\ref{sec:level3}, we give the method to calculate SKR with two-intensity decoy setups and four-intensity setups. The conclusion and a discussion are given in Sec.\ref{sec:level4}.

\section{\label{sec:level2}PASSIVE-DECOY TF-QKD PROTOCOL}

The system of our passive-decoy TF-QKD scheme is given in Fig.\ref{fig:system}. The setups on Alice have no difference from that on Bob, thus we take Alice as an example. Weak coherent pulse (WCP) is generated at Alice side, $\rm BS_1$ splits the pulses into two paths (path r and path s), the transmittance of $\rm BS_1$ is $t_1$. Path r and path s are the same in length so the pulses at these two paths will interfere in the passive-decoy setup (PDS). After interference, the outcome mode $a_1$ is sent to untrusted party Charlie. The PDS is the setup we designed to implement our passive-decoy scheme, it mainly consists of BSs and photon detectors, specifically, Fig.\ref{fig:intensity_2} and Fig.\ref{fig:intensity_4} are two PDSs with two-intensity settings and four-intensity settings. 
the specific steps of our passive-decoy TF-QKD scheme can be described as follows \cite{cui2019twin}:

{\bfseries step 1. }Alice and Bob randomly choose code mode or decoy mode in each trial. 

{\bfseries step 2.a. }When Alice (Bob) chooses code mode, The intensity modulator (IM) blocks the pulses on path s, here we choose Lithium Niobate-based Mach-Zehnder (LNMZ) intensity modulator for its advantages of low driving voltage, high bandwidth, adjustable chirp. The phase modular ($\rm PM_1$) on path r modulates the signal phase according to the random classical key bit 0 or 1. By doing these, Alice (Bob) prepares a weak coherent state $|\pm\sqrt{\mu}\rangle_A$ ($|\pm\sqrt{\mu}\rangle_B$) mentioned in Ref. \cite{cui2019twin}.

{\bfseries step 2.b. }When Alice (Bob) chooses decoy mode, $\rm PM_1$ and $\rm PM_2$ randomize the phase of the pulses on each path and leave the IM on path s unmodulated, the randomized phase will never be publicly announced. By doing these, mixed state in Fock space is generated on path r and path s, and the photon distribution of the mixed state is depending on the mode of photon detectors in PDS. 

{\bfseries step 3. } For each trial, Charlie publicly announces which detector (L or R) clicks or is a no click event.

{\bfseries step 4. } After repeating steps 1 to 3 for sufficient times, Alice and Bob publicly announce which trials are code modes and which trials are decoy modes. For the trials in which Alice and Bob both select the code mode and Charlie announces a click event, the raw key bits are generated. For the trials in which Alice and Bob both select decoy mode, Alice and Bob can estimate the yield $Y_{nm}$,  $Y_{nm}$ is the probability of Charlie announcing click event provided Alice emits an $ n$-photon state and Bob emits an $ m$-photon state in a decoy mode.

In step 2, the IM we choose is LNMZ intensity modulator. Since the only purpose of the IM is blocking the pulses on path s in code mode with no information encoded, thus it won't lead to side channel attacks. On the other hand, LNMZ intensity modulator has high bandwidth (usually GHz), making it suitable for systems with high repetition rate like ours.  Besides, the extinction ratio of LNMZ intensity modulator is usually more than 25 dB, which means the error rate caused by the IM could be neglected compared with the code rate of our system (see detailed derivation in appendix A).

In the scheme, the secret key rate per trial in a code mode is given as follow
\begin{equation}
R=Q_\mu[1-fh(er_\mu,1-er_\mu)-I^u_{AE}].
\label{equ:key}
\end{equation}
Here $h(x,y)=-xlog_2x-ylog_2y+(x+y)log_2(x+y)$, $Q_\mu$ is the counting rate in code mode, $er_\mu$ is the error rate of raw key bits, $Q_\mu$ and $er_\mu$ can be directly observed experimentally. $I^u_{AE}$ is the upper bound of the information leakage, and $I^u_{AE}$ can be solved by the following optimization problem
\[I^u_{AE}={\rm max} \, h(\frac{x_{00}}{Q_\mu},\frac{x_{10}}{Q_\mu})+h(\frac{x_{11}}{Q_\mu},\frac{x_{01}}{Q_\mu}),\]
\begin{equation}
\begin{cases}
0 \leq x_{00} \leq |\sum_{n,m=0}\sqrt{P_{2n}P_{2m}Y_{2n,2m}}|^2 ,\\
0 \leq x_{10} \leq |\sum_{n,m=0}\sqrt{P_{2n+1}P_{2m}Y_{2n+1,2m}}|^2 ,\\
0 \leq x_{11} \leq |\sum_{n,m=0}\sqrt{P_{2n+1}P_{2m+1}Y_{2n+1,2m+1}}|^2,\\
0 \leq x_{01} \leq |\sum_{n,m=0}\sqrt{P_{2n}P_{2m+1}Y_{2n,2m+1}}|^2,\\
  x_{00}+x_{10}+x_{11}+x_{01}=Q_\mu,
\end{cases}
\label{equ:case}
\end{equation}
in which $P_k=e^{-\mu}\mu^k/k!$ is Poissonian distribution. $P_n$ $(P_m)$ is determined by the input parameter $\mu$, which is chosen by the parties prior to initiating the experiment.

\section{\label{sec:level3}Calculation of the key rates}

To get the secret key rate R in code mode, we need to know the information leakage $I^u_{AE}$ in Eq.(\ref{equ:case}). For that $P_n, P_m$ and $Q_\mu$ in Eq.(\ref{equ:case}) can be observed in practical experiments, thus it is necessary to estimate $Y_{n,m}$. In our scheme, $Y_{n,m}$ can be bounded with the passive-decoy state method. The gain of the decoy states satisfies
\begin{equation}
Q_d=\sum_{n,m}Pr_a(n)Pr_b(m)Y_{n,m}.
\label{eq:decoy}
\end{equation}
Here $Q_d$ is the counting rate in decoy mode, $Pr_a(n)$ ($Pr_b(m)$) is the $n$-photon ($m$-photon) distribution of the state on Alice (Bob) in decoy mode, which depends on the mode of photon detectors on PSD in our passive-decoy scheme. In ideal case with infinite decoy states, we can list infinite linear equations like Eq.(\ref{eq:decoy}) to calculate $Y_{n,m}$ accurately. On the other hand, finite decoy states can also help to estimate the lower bound for yields and it is more feasible in practical system.

In our scheme, when the two-intensity setup is implemented in PDS as shown in Fig.\ref{fig:intensity_2}, pulses on path r and path s will interfere at $\rm BS_2$ with two outcomes $ a_1$ and $ a_2$, $ a_1$ is sent to Charlie and $ a_2$ is detected by $\rm D_1$. Denote the detect efficiency of $\rm D_1$ is $\eta$, the darkcount rate is $\eta_d$, we can get
\begin{equation}
\begin{cases}
P_{u}(n)={(1-\eta)}^n(1-\eta_d),\\
P_{c}(n)=1-P_{u}(n),
\end{cases}
\end{equation}
in which $P_{c}(n)$ and $P_{u}(n)$ are probabilities that detector click or not when $n$ photons come. With these two probabilities, we have two different conditional photon number distributions according to the mode of $\rm D_1$ (see the details in the appendix B)

\begin{equation}
Pr^1(k)=\frac{\sum\limits_{q=0}^{\infty}P{(q,k|\mu_1,\mu_2)}P_c(q)}{\sum\limits_{q,k=0}^{\infty}P{(q,k|\mu_1,\mu_2)}P_c(q)},
\label{equ:click}
\end{equation}
and
\begin{equation}
Pr^0(k)=\frac{\sum\limits_{q=0}^{\infty}P{(q,k|\mu_1,\mu_2)}P_u(q)}{\sum\limits_{q,k=0}^{\infty}P{(q,k|\mu_1,\mu_2)}P_u(q)}.
\label{equ:noclick}
\end{equation}

Here $Pr^1(k)$ is the probability of finding $k$ photons in $ a_1$ when $\rm D_1$ produces a click, $Pr^0(k)$ is the probability of that when $\rm D_1$ doesn't click, $P(q,k|\mu_1,\mu_2)$ is the conditional probability of finding $k$ photons in $ a_1$ and $q$ photons in $ a_2$ when the mean photon numbers of pulses in path s and path r are $\mu_1$ and $\mu_2$, $P(q,k|\mu_1,\mu_2)$ is derived in appendix. Then substituting $Pr^1(k)$ and $Pr^0(k)$ into Eq.(\ref{eq:decoy}), we can have:

\begin{equation}
\begin{cases}
Q_d^{0,0}=\sum_{n,m}Pr_a^0(n)Pr_b^0(m)Y_{n,m},\\
Q_d^{0,1}=\sum_{n,m}Pr_a^0(n)Pr_b^1(m)Y_{n,m},\\
Q_d^{1,0}=\sum_{n,m}Pr_a^1(n)Pr_b^0(m)Y_{n,m},\\
Q_d^{1,1}=\sum_{n,m}Pr_a^1(n)Pr_b^1(m)Y_{n,m},\\
\end{cases}
\label{equ:gain}
\end{equation}
in which $Q_d^{0,0}$, $Q_d^{0,1}$, $Q_d^{1,0}$ and $Q_d^{1,1}$ are counting rates on different detector modes in decoy mode . For example, in Eq.(\ref{equ:gain}) $Q_d^{0,1}$ is the counting rate when the detector on Alice doesn't click and the detector on Bob produces a click, $Pr^0_a(n)$ and $Pr^0_b(n)$ are defined as same as $Pr^0{(n)}$ while the subscripts $a$ and $b$ denote the distributions on Alice's and Bob's sides respectively.

In practice, after observing $Q_d^{0,0}$, $Q_d^{0,1}$, $Q_d^{1,0}$ and $Q_d^{1,1}$, we can give approximations to $Y_{0,0}$, $Y_{0,1}$, $Y_{1,0}$, $Y_{2,0}$, $Y_{0,2}$ and $Y_{2}$ by linear programming using Eq.(\ref{equ:gain}), where $Y_{2}$ denots the yields that are from decoy trials in which Alice and Bob share two photons in total, which means $ n+ m=2$. Then following the same technique in Sec.V of Ref. \cite{cui2019twin}, we substitute these yields into Eq.(\ref{equ:case}) to bound the information leakage $I^u_{AE}$ and calculate the final key rate $R$ in Eq.(\ref{equ:key}).

After presenting the scheme of generating two-intensity decoy state passively, we decide to go further, thus we add $\rm BS_3$ and two photon detectors at the end, as shown in Fig.\ref{fig:intensity_4}. There are four different modes with two photon detectors in the setup, thus we can get four different conditional photon distributions according to the modes of detector $\rm D_1$ and $\rm D_2$, which means four-intensity decoy states (see the details in the appendix B)

\begin{equation}
Pr^{00}(k) = \frac{\sum\limits_{j,l=0}^{\infty}P{(k,j,l|\mu_1,\mu_2)}P_u(j)P_u(l)}{\sum\limits_{k,j,l=0}^{\infty}P{(k,j,l|\mu_1,\mu_2)}P_u(j)P_u(l)},
\end{equation}
\begin{equation}
Pr^{10}(k) = \frac{\sum\limits_{j,l=0}^{\infty}P{(k,j,l|\mu_1,\mu_2)}P_c(j)P_u(l)}{\sum\limits_{k,j,l=0}^{\infty}P{(k,j,l|\mu_1,\mu_2)}P_c(j)P_u(l)},
\end{equation}
\begin{equation}
Pr^{01}(k) = \frac{\sum\limits_{j,l=0}^{\infty}P{(k,j,l|\mu_1,\mu_2)}P_u(j)P_c(l)}{\sum\limits_{k,j,l=0}^{\infty}P{(k,j,l|\mu_1,\mu_2)}P_u(j)P_c(l)},
\end{equation}
\begin{equation}
Pr^{11}(k) = \frac{\sum\limits_{j,l=0}^{\infty}P{(k,j,l|\mu_1,\mu_2)}P_c(j)P_c(l)}{\sum\limits_{k,j,l=0}^{\infty}P{(k,j,l|\mu_1,\mu_2)}P_c(j)P_c(l)},
\end{equation}

in which $Pr^{00}(k)$, $Pr^{10}(k)$, $Pr^{01}(k)$ and $Pr^{11}(k)$ are the photon distributions in $ a_1$ given that $\rm D_1$ and $\rm D_2$ are in different modes. For example, $Pr^{10}(k)$ is the probability of finding $k$ photons in $\rm a_1$ when $\rm D_1$ produces a click and $\rm D_2$ doesn't. Note that when the transmittance of $\rm BS_3$ is $t_3=0.5$, we will get $Pr^{10}(k)=Pr^{01}(k)$, which means $Pr^{10}(k)$ and $Pr^{01}(k)$ are not independent. To get more useful constraints in linear program, we let $0.5<t_3<1$. For readability we write $Pr^{00}$, $Pr^{10}$, $Pr^{01}$, $Pr^{11}$ as $Pr^{1}$, $Pr^{2}$, $Pr^{3}$, $Pr^{4}$, and define the following gains: $Q_d^{4,4}$, $Q_d^{1,4}$, $Q_d^{2,4}$, $Q_d^{3,4}$, $Q_d^{4,1}$, $Q_d^{4,2}$, $Q_d^{4,3}$, $Q_d^{1,1}$, $Q_d^{2,2}$ and $Q_d^{3,3}$. For example, $Q_d^{1,4}$ is the counting rate when both $\rm D_1$ and $\rm D_2$ on Alice don't click and the both $\rm D_1$ and $\rm D_2$ on Bob produce click, then like what we do in two-intensity cases, after observing these gains, we can give approximations to $Y_{0,0}$, $Y_{0,1}$, $Y_{1,0}$, $Y_{2,0}$, $Y_{0,2}$ and $Y_{2}$ by linear programming, and we can get the final key rate $R$ using Eq.(\ref{equ:key},\ref{equ:case}). This’s the passive scheme of generating four-intensity decoy state with two BS and two photon detectors.

Comparing Fig.\ref{fig:intensity_2} with Fig.\ref{fig:intensity_4}, we can know that Alice could increase the mode of the detectors by adding a BS at the end to get more constrains in linear programing and improve key rates. Following this idea, we can actually generalize our setup to infinite passive-decoy case. we denote the joint photon number distribution of the passive setup with $n$ ($n$=1,2,3...) intensity settings is $p^n$, we can get a recursive formula
\begin{equation}
p^n=p^{n-1}p{(n,m|\mu_1,\mu_2)}
\label{equ:n_case}
\end{equation}

From Eq.(\ref{equ:n_case}), we can get the conditional probabilities of photon number distribution in $ a_1$ on each mode of the detectors, one may refer to  the appendix for detailed derivations. 

Now we carry out a numerical simulation to evaluate the performance of our two-intensity and four-intensity protocols. We focus on the symmetric case, which means that Charlie is at the middle of Alice and Bob and all other device parameters on Alice’s side and Bob’s side are identical. The simulation model can be found in the appendix of Ref.\cite{cui2019twin}, the parameters in our simulation are listed in Table.1.

\begin{table}[H]
\caption{\label{tab:table1}%
Experimental parameters used in the numerical simulation. $\eta$ is the detect efficiency of single photon detectors, $\eta_d$ is darkcount rate, $f$ is the correction efficiency, $e_d$ is the misalignment error probability. }
\begin{ruledtabular}
\begin{tabular}{ccccc}
$\eta$&$\eta_d$&
\multicolumn{1}{c}{\textrm{fiber loss}}&
\multicolumn{1}{c}{\textrm{$f$}}&
\multicolumn{1}{c}{\textrm{$e_d$}}\\
%\mbox{Three}&\mbox{Four}&\mbox{Five}\\
\hline
20\%&$1*10^{-7}$&\mbox{0.2 dB/km}&\mbox{1.15}&\mbox{0.03}\\
\end{tabular}
\end{ruledtabular}
\end{table}

Our results are shown in Fig.\ref{fig:result}. It shows that, in two-decoy settings, there is a gap between active decoy scheme and passive-decoy scheme, while in four-decoy settings, our passive-decoy scheme performs as good as the active one and even slightly better at a long distance. This shows that our passive-decoy state method can perform as good as the active one. Furthermore, our method can eliminate side channels on sources which may be caused by active modulation. Thus our scheme is very promising for applications of practical systems.

Besides, we also add a simulation of the passive setup with eight decoys, and compare it with the infinite decoy states.The result shows that the secret key rate of passive setup with the eight decoys surpasses that of the four decoys at a long range of distance. More importantly, the gap between the passive setup with the eight decoys and the infinite decoy states is narrow, which shows our protocol can present satisfactory performance with just a few 
detectors.

\begin{figure}%[H]
 {\centering
 \includegraphics[width = .5\textwidth]{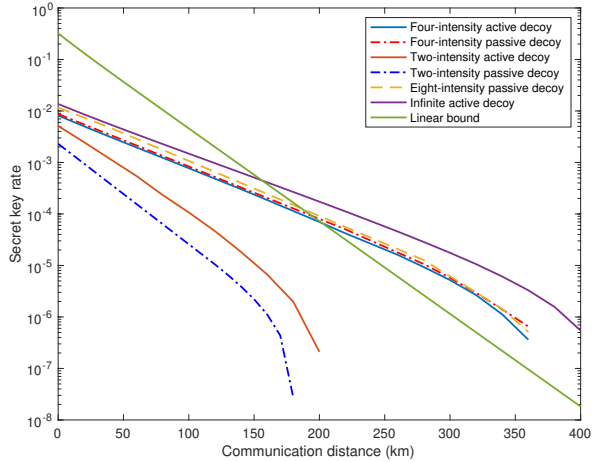}
 }
 \caption{Secret key rate in logarithmic scale of two-intensity, four-intensity, eight-intensity in passive decoy scheme and two-intensity, four-intensity in active decoy scheme. The infinite case in active decoy scheme and linear key rate bound are also shown in the figure.  }
 \label{fig:result}
 \end{figure}

\section{\label{sec:level4}Conclusion And Discussion}
\vspace{-3mm}
We have proposed the scheme of passive-decoy state TF-QKD, in which different decoy states are generated passively according to the modes of photon detectors. In this way, our scheme can eliminate side channel on source caused by actively modulating of the source intensity. Moreover we presented a simulation comparing the key generation rate with that in active-decoy, it shows our result can performs as good as that in active decoy TF-QKD, and our scheme could reach satisfactory secret key rates with just a few photon detectors. Thus our work provides a useful approach to improve the performance of TF-QKD in practical implementation.

Nevertheless, although our scheme can remove certain side channel caused by actively modulating intensity of the source,  we can not remove all side channels, there are still further work to do. For example, the scheme is still threatened by Trojan horse attack although it's not easy in practice, and an imperfect real-life source may emit imperfect states with varied frequency spectra, which will also lead to side channel. Fortunately, imperfect states preparation may be remedied by Ref. \cite{wang2019practical}, which proposed that an imperfect source $S$ and a perfect source $P$ are identical in side-channel space if there exists a quantum process that can map the $S$ to $P$. Besides, the intensities of the photon pulses could be unstable. To solve this issue, Refs. \cite{wang2009decoy,hu2010reexamination} have discussed these issues. Ref. \cite{wang2009decoy} bounded the fraction of single-photon counts in the decoy-state QKD, while Ref. \cite{hu2010reexamination} proposed a theory that decoy-state QKD with an unstable source can work as if the source were stable provided the condition:
$
\max \limits_j(\frac{p_ja_{kj}}{p'_ja'_{kj}}) \leq \max \limits_j(\frac{p_ja_{2j}}{p'_ja'_{2j}}) < \max \limits_j(\frac{p_ja_{1j}}{p'_ja'_{1j}})$, for all $k \geq 2
$
hold. In our scheme, this condition still holds by optimizing the transmittance of BS(s), therefore, our scheme can work under unstable source in principle. Moreover, there are still side channels in our scheme due to bases switching, i.e., different values in phase modulation in the code mode. It would be interesting to study this issue in the future.

\section*{Acknowledgements}
This work has been supported by the National Key Research and Development Program of China (Grant No. 2016YFA0302600), the National Natural Science Foundation of China (Grant Nos. 61822115,61961136004,
61775207, 61702469, 61771439, 61627820,61675189), National Cryptography Development Fund (Grant No. MMJJ20170120) and Anhui Initiative in Quantum Information Technologies.

\section*{appendix A:THE AFFECTION OF IM IN THE SYSTEM}
\appendix
\setcounter{equation}{0}
\renewcommand\theequation{\arabic{equation}}
We take Alice as an example,  for simplicity, we let the extinction ratio of the IM be 20 dB, the intensity of the weak coherent pulses from laser be $\mu$, the transmittance of $\rm BS_1$ is $t_1$, thus we have 
\begin{equation}
\begin{cases}
\mu_s = (1-t_1)\mu,\\
\mu_r = t_1\mu,
\end{cases}
\end{equation}
in which $\mu_s$, $\mu_r$ are the intensity of the pulses at the two outputs of $\rm BS_1$, then pulses on path s will be blocked by the IM, the intensity of the output pulses is $\mu_s'=\mu_s/100$, thus we can get the probability 
\begin{equation}
\begin{cases}
P(\mu_s',n=0) = e^{-\mu_s'},\\
P(\mu_s',n\neq0) = 1-e^{-\mu_s'}.\\
\end{cases}
\end{equation}

Where $P(n=0,\mu_s')$ is the probability of finding no photons from the output of path s after blocking of the IM, $P(n\neq0,\mu_s') = 1-e^{-\mu_s'}$ is the probability of finding at least one photon from the output of path s.

For simplicity, we take the error rate $er=0.5$ on $a_1$ as long as there are photons come from path s, thus we have the total error rate 
\begin{equation}
er_{total}=0.5P(\mu_s',n\neq0)=0.5(1-e^{-\mu_s'}).
\end{equation}

Since $\mu_s'=\mu_s/100$, thus $er_{total}$ is a small value, we take $\mu_s=0.1$ for example, we get $er_{total}=0.00005$, it's a small value that it won't affect the performance of our system.

\section*{appendix B:passively generate different photon number statistics}
\appendix
\setcounter{equation}{0}
\renewcommand\theequation{\arabic{equation}}

Here we focus on the PDS in the whole passive-decoy TF-QKD system, as shown in Fig.\ref{passive_part}. Let us first consider the interference with two pure coherent states ${|n\rangle}={a^\dag}^n|0\rangle$ and ${|m\rangle}={b^\dag}^m|0\rangle$, where $a^\dag$ and $b^\dag$ are creation operators, $n$ and $m$ are photon numbers. For $a^\dag$ and $b^\dag$ at a BS:
\begin{equation}
\begin{cases}
a^\dag =\sqrt{t}c^\dag+\sqrt{1-t}d^\dag,\\
b^\dag =\sqrt{1-t}c^\dag-\sqrt{t}d^\dag,
\end{cases}
\end{equation}

Here $t$ is the transmittance of the $\rm BS_2$, $c^\dag$ and $d^\dag$ are creation operators of the quantum state of output $ a_1$ and $a_2$. Denote $U_{BS_2}$ is the unitary operator of $\rm BS_2$, thus we have 
\begin{equation}
\begin{aligned}
&{U_{BS_2}|nm\rangle}\\
&={(\sqrt{t}c^\dag+\sqrt{1-t}d^\dag)}^n{(\sqrt{1-t}c^\dag-\sqrt{t}d^\dag)}^m|00\rangle.
\label{equ:BS1}
\end{aligned}
\end{equation}
from Eq.(\ref{equ:BS1}) we can get the probability of finding $k$ photons at mode $ a_1$, it's a problem of statistics.
%\begin{equation}
\[
\begin{aligned}
&P{(k|n,m)}\\
&=\sum_{\max(0,k-m)}^{\min(k,n)}C_n^iC_m^{(k-i)}\frac{\sqrt{t^n{(1-t)}^m}}{\sqrt{m!n!}}\sqrt{k!(n+m-x)!}.
\end{aligned}
%\end{equation}
\]
Here $P{(k|n,m)}$ is the probability of finding $k$ photons on $ a_1$ given the coherent states on path s and path r containing $n$ and $m$ photons respectively. 
With $P{(k|n,m)}$ we can then get the joint probability $P{(k,n,m|\mu_1,\mu_2)}$
\begin{equation}
P{(k,n,m|\mu_1,\mu_2)}= { P}{(\mu_1,n)}{ P}{(\mu_2,m)}P{(k|n,m)}, %算出概率
\end{equation}
where $P{(k,n,m|\mu_1,\mu_2)}$ means the pulses in path r, s are phase-randomized weak coherent states with mean photon number $\mu_1$ and $\mu_2$, ${ P}(\mu,n)=e^{-\mu}\mu^n/n!$ is Poissonian distribution. Summing up $n$ and $m$ and we can get
\begin{equation}
P{(k|\mu_1,\mu_2)}= \sum_{n,m=0}{ P}{(\mu_1,n)}{ P}{(\mu_2,m)}P{(k|n,m)},
\label{equ:find_k}
\end{equation}
where $P{(k|\mu_1,\mu_2)}$ is the photon distribution on mode $ a_1$ given two phase-randomized weak coherent states interfere at $\rm BS_2$. In Fig.\ref{passive_part}, the photon distribution on mode $ a_2$ is the same as mode $ a_1$.
\begin{equation}
P{(q|\mu_1,\mu_2)}= \sum_{n,m=0}{ P}{(\mu_1,n)}{ P}{(\mu_2,m)}P{(q|n,m)},
\label{equ:q_no_a2}
\end{equation}
thus we can get $P{(k,q|\mu_1,\mu_2)}$ 
\begin{equation}
P{(k,q|\mu_1,\mu_2)}= P{(k|\mu_1,\mu_2)}P{(q|\mu_1,\mu_2)}.
\end{equation}
Here $P{(k,q|\mu_1,\mu_2)}$ is the joint probability of finding $k$ photons on mode $ a_1$ and $q$ photons on mode $ a_2$. Denote $\eta$ is detect efficiency, $\eta_d$ is darkcount rate of photon detector $\rm D_1$, we have
\[\begin{cases}
P_u(n)={(1-\eta)}^n(1-\eta_d),\\
P_c(n)=1-P_u(n),
\end{cases}\]
in which $P_c(n)$ and $P_u(n)$ are probabilities that the detector $\rm D_1$ produces a click and no click when $n$ photons come, thus there are two modes of the detector $\rm D_1$ according to its click event and no click event. consequently, there are two different conditional photon distribution on $a_1$ according to the mode of $\rm D_1$
\begin{equation}
Pr^1(k)=\frac{\sum\limits_{q=0}^{\infty}P{(q,k|\mu_1,\mu_2)}P_{c}(q)}{\sum\limits_{q,k=0}^{\infty}P{(q,k|\mu_1,\mu_2)}P_{c}(q)},
\end{equation}
\begin{equation}
Pr^0(k)=\frac{\sum\limits_{q=0}^{\infty}P{(q,k|\mu_1,\mu_2)}P_{u}(q)}{\sum\limits_{q,k=0}^{\infty}P{(q,k|\mu_1,\mu_2)}P_{u}(q)},
\end{equation}
where the subscripts '1' and '0' of $Pr^1(k)$ and $Pr^0(k)$ means the detector clicks or not when pulses come.

\begin{figure}[hbt]
\appendix      
\setcounter{figure}{0} 
{ \centering\includegraphics[width = .25\textwidth]{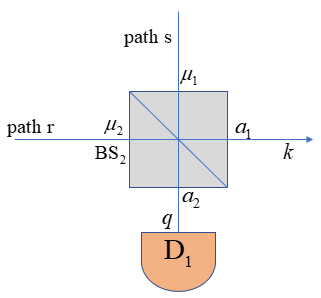}

}
\caption{Passive-decoy state setup with two intensity settings }
\label{passive_part}
\end{figure}

The derivation above clarifies our passive setup with two-intensity settings, and we also investigate the four-intensity settings, the approach is to add a $\rm BS_3$ at the end and two detectors for the outcome signals as shown in Fig.\ref{two_intensity_passive}. Thus there are four different conditional photon distributions on $ a_1$, $Pr^{00}(y)$, $Pr^{01}(y)$, $Pr^{10}(y)$, $Pr^{11}(y)$, according to the mode of the two detectors, We have get the probability of finding $q$ photons on $ a_2$ in Eq.(\ref{equ:q_no_a2}), denote the transmittance of $\rm BS_3$ is $ t_3$, we can get the probability of finding $j$, $ l$ photons at the two output of $\rm BS_3$ with $q$ photons be splitted at $\rm BS_3$
\begin{equation}
P{(j,l|q)} = C_q^{j}t_3^{l}(1-t_3)^{l},
\label{equ:BS}
\end{equation}
thus we can get 
\begin{equation}
P{(j,l|\mu_1,\mu_2)} = \sum\limits_{q=0}P{(j,l|q)}P{(q|\mu_1,\mu_2)}.
\end{equation}
Here $P{(j,l|\mu_1,\mu_2)}$ is the probability of finding $j$ photons on $\rm D_1$ and $l$ photons on $\rm D_2$, when there are two weak coherent states with the mean photon number of $\mu_1$ and $\mu_2$ on path s and path r.
Then we have the joint probability 
\begin{equation}
P{(k,j,l|\mu_1,\mu_2)} = P{(j,l|\mu_1,\mu_2)}P{(k|\mu_1,\mu_2)}.
\end{equation}
Here $P{(k,j,l|\mu_1,\mu_2)}$ is the joint probability of finding $k$ on $ a_1$, $ j$ photons on $\rm D_1$ and $ l$ photons on $\rm D_2$, then we can get the conditional probabilities of photon number distribution on $ a_1$ according to the mode of the detectors $\rm D_1$ and $\rm D_2$

\begin{equation}
Pr^{00}(k) = \frac{\sum\limits_{j,l=0}^{\infty}P{(k,j,l|\mu_1,\mu_2)}P_{u}(j)P_{u}(l)}{\sum\limits_{k,j,l=0}^{\infty}P{(k,j,l|\mu_1,\mu_2)}P_{u}(j)P_{u}(l)},
\end{equation}

\begin{equation}
Pr^{10}(k) = \frac{\sum\limits_{j,l=0}^{\infty}P{(k,j,l|\mu_1,\mu_2)}P_{c}(j)P_{u}(l)}{\sum\limits_{k,j,l=0}^{\infty}P{(k,j,l|\mu_1,\mu_2)}P_{c}(d_1)P_{u}(l)},
\end{equation}

\begin{equation}
Pr^{01}(k) = \frac{\sum\limits_{j,l=0}^{\infty}P{(k,j,l|\mu_1,\mu_2)}P_{u}(j)P_{c}(l)}{\sum\limits_{k,j,l=0}^{\infty}P{(k,j,l|\mu_1,\mu_2)}P_{u}(j)P_{c}(l)},
\end{equation}

\begin{equation}
Pr^{11}(k) = \frac{\sum\limits_{j,l=0}^{\infty}P{(k,j,l|\mu_1,\mu_2)}P_{c}(j)P_{c}(l)}{\sum\limits_{k,j,l=0}^{\infty}P{(k,j,l|\mu_1,\mu_2)}P_{c}(d_1)P_{c}(l)}.
\end{equation}

\begin{figure}[hbt]
 {\centering\includegraphics[width = .25\textwidth]{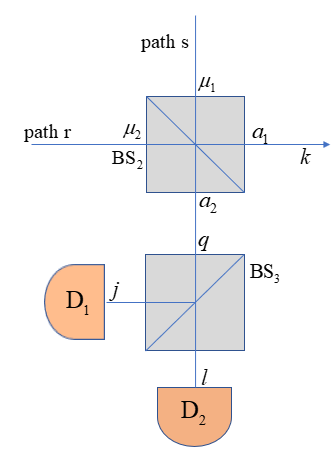}

}
\caption{Passive-decoy state setup with four intensity settings}
\label{two_intensity_passive}
\end{figure}

The derivation above clarifies our passive-decoy setup with four-intensity settings, the main idea is adding a $\rm BS$ at the end of the setup with two-intensity settings and two detectors detecting the output pulses. Thus the modes of detectors have increased from two to four, which means we increased the number of passive decoy intensities from two to four. 

Following this idea, we can generalize the setup to infinite passive-decoy case. First, we denote the joint probability of the passive-decoy setup with $n-1$ ($n$=1,2,3...) intensity settings is $P{(k,d_1,d_2...d_{n-1}|\mu_1,\mu_2))}$, which means we add $n-1$ photon detectors at the end and the joint probability of finding $k$ photons on mode $a_1$ and finding $d_n$ photons on the corresponding photon detector ${\rm D}_n$ is $P{(k,d_1,d_2...d_{n-1}|\mu_1,\mu_2)}$. Then if we add another BS at the end and two photon detectors at the output mode, the setup becomes a setup with $n$ intensity settings, this means in the setup with $n-1$ settings, the pulse on the ${\rm D}_{n-1}$ is splitted by the added BS and detected by the add photons detectors, denote the transmittance of this BS is $t$, we can get
\begin{equation}
P{(n,m|d_{n-1})}=C_q^{n}t^{m}(1-t)^{m},
\end{equation}
this is the same as given in Eq.(\ref{equ:BS}), and in $P{(k,d_1,d_2...d_{n-1}|\mu_1,\mu_2)}$, we have 
\begin{equation}
P{(d_{n-1}|\mu_1,\mu_2)}=\sum\limits_{k,d_1,d_2...d_{n-2}=0}P{(k,d_1,d_2...d_{n-1}|\mu_1,\mu_2)}.
\end{equation}
Here $P{(d_{n-1}|\mu_1,\mu_2)}$ is the probability of finding $ d_{n-1}$ photons on detector ${\rm D}_{n-1}$ in the setup with $n-1$ intensity settings. Thus we can get
\begin{equation}
P{(n,m|\mu_1,\mu_2)}=\sum\limits_{d_{n-1}=0}P{(n,m|d_{n-1})}P{(d_{n-1}|\mu_1,\mu_2)}.
\label{eq:conditional_prob}
\end{equation}
Here $P{(n,m|\mu_1,\mu_2)}$ is the probability of finding n and m photons on the two output mode given two phase randomized weak coherent states on path s and path r, and the mean photon number of these two phase randomized WCP is $\mu_1$ and $\mu_2$

With the probability on Eq.(\ref{eq:conditional_prob}), we can get the joint probability of the passive setup with $n$ ($n$=1,2,3...) intensity settings 
\begin{equation}
\begin{aligned}
&P{(k,d_1...d_{n-2},n,m|\mu_1,\mu_2)}\\
&=P{(k,d_1...d_{n-1}|\mu_1,\mu_2)}P{(n,m|\mu_1,\mu_2)}.
\end{aligned}
\end{equation}
For readability, we substitute $n$, $m$ with $d_{n-1}$ and $d_n$, thus $P{(k,d_1...d_{n-2},d_{n-1},d_n|\mu_1,\mu_2))}=P{(k,d_1...d_{n-2},n,m|\mu_1,\mu_2)}$. And we denote $P^n=P{(k,d_1...d_{n-2},d_{n-1},d_n|\mu_1,\mu_2)}$ is the joint probability on passive setup with n intensity settings, $P^{n-1}=P{(k,d_1...d_{n-1},n,m|\mu_1,\mu_2)}$ is the joint probability on passive setup with $n-1$ intensity settings, thus we have 
\begin{equation}
P^n=P^{n-1}P{(n,m|\mu_1,\mu_2)}.
\label{eq:intensity_n}
\end{equation}
Eq.(\ref{eq:intensity_n}) is a recursive formula. 
In the setup with $n$ intensity settings, there are $n$ photon detectors, thus there are ${\rm M}=\sum\limits_{m=0}^nC_n^m$ modes according to whether the detectors click or not, which means there are ${\rm M}=\sum\limits_{m=0}^nC_n^m$ conditional probabilities of photon number distribution on $ a_1$ on each mode of detectors. We take $ Pr_{100...000}(k)$ in Eq.(\ref{equ:P_{100...000}(k)}) for example

\begin{widetext}
\begin{equation}
Pr_{100...000}(k) = \frac{\sum\limits_{d_1,d_2...d_n=0}^{\infty}P{(k,d_1,...,d_n|\mu_1,\mu_2)}P_{c}(d_1)P_{u}(d_2,...,d_n)}{\sum\limits_{k,d_1,d_2,...,d_n=0}^{\infty}P{(k,d_1,...,d_n|\mu_1,\mu_2)}P_{c}(d_1)P_{u}(d_2,..,d_n)}.
\label{equ:P_{100...000}(k)}
\end{equation}
\end{widetext}

$Pr_{100...000}(k)$ is the conditional probability of photon number distribution on $ a_1$ in the mode that only the detector $\rm D_1$ clicks and the rest don't among the $ n$ detectors added. For other modes of the detectors, the conditional probability of photon number distribution is similar to Eq.(\ref{equ:P_{100...000}(k)}).

\nocite{*}
\bibliography{TFQKD}% Produces the bibliography via BibTeX.

\end{document}